%% file: main.tex
\documentclass[runningheads, inline]{llncs}
\usepackage[draft]{common}

\usepackage{graphics}
\usepackage{float}
\usepackage{graphicx}
\usepackage{tabularx}
\usepackage{amsmath}
\usepackage{amsfonts}
\usepackage{url}
\usepackage{booktabs,caption}
\usepackage{longtable}
\usepackage{multirow}
\usepackage{makecell}
\usepackage{hyperref}
\usepackage{xcolor}
\usepackage{subcaption}
\usepackage{xspace}
\usepackage{moresize}

\hypersetup{colorlinks=true,
	linkcolor=blue,
	citecolor=blue,
	urlcolor=blue}

\usepackage[autolanguage]{numprint}

\author{%
  Daniel Perez\inst{1} \and%
  Sam M. Werner\inst{1} \and%
  Jiahua Xu\inst{2,4} \and%
  Benjamin Livshits\inst{1,2,3}
}
\authorrunning{D. Perez, S.M. Werner, J. Xu and B. Livshits}
\institute{%
  Imperial College London \and%
  University College London, Centre for Blockchain Technologies \and%
  Brave Software \and
  École polytechnique fédérale de Lausanne
}

\definecolor{light-gray}{gray}{0.95}

\newcommand{\coin}[1]{\texttt{#1}}
\newcommand{\contractaddr}[2][\small]{{#1\href{https://etherscan.io/address/#2}{\texttt{#2}}}}

\newcommand{\StartDate}{May~7,~2019\xspace}
\newcommand{\EndDate}{September~6,~2020\xspace}

\newcommand{\linktoonline}[1]{\href{https://fc21.ifca.ai/papers/144.pdf}{#1}} 

\newcommand{\onlineversion}[1]{#1} 
\newcommand{\printversion}[1]{} 

\title{Liquidations: DeFi on a Knife-edge}

\begin{document}
\maketitle
\input{sections/0_abstract}

\input{sections/1_introduction}

\input{sections/2_background}

\input{sections/3_plfs}

\input{sections/4_methodology}

\input{sections/5_analysis}

\input{sections/7_discussion}

\input{sections/8_related_work}

\input{sections/9_conclusion}

\bibliographystyle{splncs04}
\bibliography{references, referencesadd}

\printversion{%
\section*{Appendix}
Appendix is available online at \linktoonline{https://fc21.ifca.ai/papers/144.pdf}.
}
\clearpage
\onlineversion{%
\appendix
\input{sections/a_contracts}
\input{sections/b_extras}

}

\end{document}

%% file: sections/0_abstract.tex
\begin{abstract}
The trustless nature of permissionless blockchains renders overcollateralization a key safety component relied upon by decentralized finance (DeFi) protocols.
Nonetheless, factors such as price volatility may undermine this mechanism. 
In order to protect protocols from suffering losses, undercollateralized positions can be \textit{liquidated}.
In this paper, we present the first in-depth empirical analysis of liquidations on protocols for loanable funds (PLFs).
We examine Compound, one of the most widely used PLFs, for a period starting from its conception to September 2020.
We analyze participants' behavior and risk-appetite in particular, to elucidate recent developments in the dynamics of the protocol.
Furthermore, we assess how this has changed with a modification in Compound's incentive structure and show that variations of only~3\% in an asset's dollar price can result in over 10m USD becoming liquidable.
To further understand the implications of this, we investigate the efficiency of liquidators.
We find that liquidators' efficiency has improved significantly over time, with currently over~70\% of liquidable positions being immediately liquidated.
Lastly, we provide a discussion on how a false sense of security fostered by a misconception of the stability of non-custodial stablecoins, increases the overall liquidation risk faced by Compound participants.
\end{abstract}

%% file: sections/1_introduction.tex
\section{Introduction}
\label{sec:introduction}

Decentralized Finance (DeFi) refers to a peer-to-peer, permissionless blockchain-based ecosystem that utilizes the integrity of smart contracts for the advancement and disintermediation of traditional financial primitives \cite{Werner2021}. 
One of the most prominent DeFi applications on the Ethereum blockchain~\cite{wood2014ethereum} are protocols for loanable funds (PLFs)~\cite{Gudgeon2020PLF}.
On PLFs, markets for loanable funds are established via smart contracts that facilitate borrowing and lending \cite{Xu2022}.
In the absence of strong identities on Ethereum, creditor protection tends to be ensured through overcollateralization, whereby a borrower must provide collateral worth more than the value of the borrowed amount.
In the case where the value of the collateral-to-borrow ratio drops below some liquidation threshold, a borrower defaults on his position and the supplied collateral is sold off at a discount to cover the debt in a process referred to as \textit{liquidation}.
However, little is known about the behavior of agents towards liquidation risk on a PLF.
Furthermore, despite liquidators playing a critical role in the DeFi ecosystem, the efficiency with which they liquidate positions has not yet been thoroughly analyzed.

In this paper, we first lay out a framework for quantifying the state of a generic PLF and its markets over time. 
We subsequently instantiate this framework to all markets on Compound~\cite{Leshner2018}, one of the largest PLFs in terms of locked funds.
We analyze how liquidation risk has changed over time, specifically after the launch of Compound's governance token.
Furthermore, we seek to quantify this liquidation risk through a price sensitivity analysis.
In a discussion, we elaborate on how the interdependence of different DeFi protocols can result in agent behavior undermining the assumptions of the protocols' incentive structures.

\point{Contributions} This paper makes the following contributions:
\begin{itemize}
    \item We present an abstract framework to reason about the state of PLFs.
    
    \item We provide an open-source implementation\footnote{\url{https://github.com/backdfund/analyzer}} of the proposed framework for Compound, one of the largest PLFs in terms of total locked funds.
    
    \item We perform an empirical analysis on the historical data for Compound, from \StartDate to \EndDate and make the following observations:
    \begin{enumerate*}[label={(\roman*)}]
    \item despite increases in the number of suppliers and borrowers, the total funds locked are mostly accounted for by a small subset of participants; 
    \item the introduction of Compound's governance token had protocol-wide implications as liquidation risk increased in consequence of higher risk-seeking behavior of participants;
    \item liquidators became significantly more efficient over time, liquidating over 70\% of liquidable positions instantly.
    \end{enumerate*}
    
    \item Using our findings, we demonstrate how interaction between protocols' incentive structures can directly result in unexpected risks to participants.
\end{itemize}

%% file: sections/2_background.tex
\section{Background}
\label{sec:background}
In this section we introduce preliminary concepts about blockchains and smart contracts necessary to the understanding of the rest of the paper.

\subsection{Blockchain}

A blockchain, such as Bitcoin~\cite{Nakamoto} or Ethereum~\cite{wood2014ethereum}, is in essence a decentralized append-only database.
Data is added to the blockchain in the form of transactions that are grouped in blocks.
Some rules are enforced by the protocols on both transactions and blocks to ensure its correct working.
Blockchains need to be able to maintain consensus of which blocks are included.
Both Bitcoin and Ethereum use the Proof-of-Work consensus that requires block producers, often called \emph{miners}, to solve a computationally expensive puzzle to produce a new block \cite{Perez2020c}.
An important point to note is that miners are allowed to choose which transactions to include in a block and in which order to include them.
This can potentially allow miners to profit from having a transaction included before another one.
This is commonly referred to as \emph{miner-extractable value}~\cite{daian2020flash}.

\subsection{Smart Contracts}
\point{Ethereum smart contracts}
On Ethereum, smart contracts are programs written in a Turing-complete language, typically Solidity~\cite{docs:solidity}, that define a set of rules that may be invoked by any network participant. 
These programs rely on the Ethereum Virtual Machine (EVM), a low-level stack machine which executes the compiled EVM bytecode of a smart contract~\cite{wood2014ethereum}.
Each instruction has a fee measured in so-called \textit{gas}, and the total gas cost of a transaction is a fixed base fee plus the sum of all instructions' gas~\cite{albert2020gasol,Perez2020f}.
The sender of a transaction must then set a gas price, the amount of \coin{ETH} he is willing to pay per unit of gas consumed for executing the transaction.
The transaction fee is thus given by the gas price multiplied with the gas cost~\cite{werner2020step,pierro2019influence}.
Within a transaction, smart contracts can store data in logs, which are metadata specially indexed as part of the transaction.
This metadata, commonly referred to as \textit{events}, is typically used to allow users to monitor the activity of a contract externally.

\point{Oracles}
One of the major challenges smart contracts face concerns access to off-chain information, i.e. data that does not natively exist on-chain.
Oracles are data feeds into smart contracts and provide a mechanism for accessing off-chain information through some third party.
In DeFi, oracles are commonly used for price feed data to determine the real-time price of assets.
For instance, via the Compound Open Price Feed~\cite{web:compoundfinance_prices}, vetted third party reporters sign off on price data using a known public key, where the resulting feed can be relied upon by smart contracts.

\point{Stablecoins}
An alternative to volatile cryptoassets is given by stablecoins, which are priced against a peg and can be either custodial or non-custodial.
For custodial stablecoins (e.g. \coin{USDC}~\cite{web:usdc}), tokens represent a claim of some off-chain reserve asset, such as fiat currency, which has been entrusted to a custodian.
Non-custodial stablecoins (e.g. \coin{DAI}~\cite{whitepaper:maker}) seek to establish price stability via economic mechanisms specified by smart contracts.
For a thorough discussion on stablecoin design, we direct the reader to \cite{Klages-Mundt2020}.

%% file: sections/3_plfs.tex
\section{Protocols for Loanable Funds (PLF)}
\label{sec:plfs}
In this section, we introduce several concepts of Protocols for Loanable Funds (PLFs) necessary for understanding how liquidations function in DeFi on Ethereum.

\subsection{Supplying and borrowing in DeFi}
In DeFi, asset supplying and borrowing is achieved via so-called \textit{protocols for loanable funds} (PLFs)~\cite{Gudgeon2020PLF}, where smart contracts act as trustless intermediaries of loanable funds between suppliers and borrowers in markets of different assets.
Unlike traditional peer-to-peer lending, deposits are pooled and instantly available to borrowers.
On a DeFi protocol, the aggregate of tokens that the PLF smart contracts hold, which equals the difference between supplied funds and borrowed funds, is termed \text{locked funds}~\cite{DeFiPulse2020}.

\subsection{Interest model}
Borrowers are charged interest on the debt at a floating rate determined by a market's underlying interest rate model.
A small fraction of the paid interest is allocated to a pool of reserves, which is set aside in case of market illiquidity, while the remainder is paid out to suppliers of loanable funds.
Interest in a given market is generally accrued through market-specific, interest-bearing derivative tokens that appreciate against the underlying asset over time.
Hence, a supplier of funds receives derivative tokens in exchange for supplied liquidity, representing his share in the total value of the liquidity pool for the underlying asset.
The most prominent PLFs are Compound~\cite{web:compoundfinance} and Aave~\cite{web:aave}, with 2.5bn USD and 2.7bn USD in total funds locked respectively, at the time of writing~\cite{DeFiPulse2020}.

\subsection{Collateralization}
Given the pseudonymity of agents in Ethereum, borrow positions need to be overcollateralized to reduce the default risk.
Thereby, the borrower of an asset is required to supply collateral, where the total value of the supplied collateral exceeds the total value of the borrowed asset.
Each asset is associated with a collateralization ratio, namely the minimum collateral-to-borrow ratio when the asset is used to collateralize a new borrow position.
For example, in order to borrow 100 USD worth of \coin{DAI} with \coin{ETH} as collateral at a collateralization ratio of 125\%, a borrower would have to lock 125 USD worth of \coin{ETH} to collateralize the borrow position.
Thus, the protocol limits monetary risk from defaulted borrow positions, as the underlying collateral of a defaulted position can be sold off to recover the debt.
The inverse of the collateralization ratio is referred to as the \textit{collateral factor}, which is the amount of a deposit that may be used as collateral.
For example, if the collateralization ratio on a PLF for the market of \coin{DAI} is $125$\%, the collateral factor would be $0.8$, implying that for each \$$1$ deposit of \coin{DAI}, the supplier may borrow \$$0.8$ worth of some other asset.

\subsection{Liquidation}
The process of selling a borrower's collateral to recover the debt value upon default is referred to as \textit{liquidation}.
A borrow position can be liquidated once the value of the collateral falls below some pre-determined liquidation threshold, i.e. the minimum acceptable collateral-to-borrow ratio.
Any network participant may liquidate these positions by paying the debt asset to acquire the underlying collateral at a discount.
Hence, liquidators are incentivized to actively monitor others' collateral-to-borrow ratios. 
Note that in practice, the amount of liquidable collateral that a single liquidator can purchase may be capped.  

\subsection{Leveraging}
In finance, leverage refers to borrowed funds being used as the funding source for additional, typically more risky capital.
In DeFi, leverage is the fundamental component of PLFs, as a borrower is required to first take up the role of a supplier and deposit funds which are to be used as leverage for his borrow positions, as we have just seen.
The typical aim of leveraging is to generate higher returns through increased exposure to a particular investment.
For example, a borrower wanting to gain increased exposure to \coin{ETH} may:
\begin{enumerate}
\item Supply \coin{ETH} on a PLF.
\item Leverage the deposited \coin{ETH} to borrow \coin{DAI}.
\item Sell the purchased \coin{DAI} for \coin{ETH}.
\item Repeat steps $1$ to $3$ as desired. 
\end{enumerate}

This behavior essentially enables users to construct so-called \textit{leveraging spirals}, whereby a user repeatedly re-supplies borrowed funds in order to get increased exposure to some cryptoasset. 
However, increased exposure comes at the cost of higher downside risk, i.e., the risk of the value of the leveraged asset or borrowed asset to decrease due to changing market conditions.

\subsection{Use Cases of PLFs}
We present the different incentives\footnote{Note that leverage on a PLF in DeFi may in part be motivated by tax benefits, as certain jurisdictions may not tax capital gains on borrowed funds. 
However, a detailed analysis of this lies outside the scope of this paper.} an agent may have for borrowing from and/or supplying to a PLF:
\begin{description}
    \item[Interest] Suppliers of funds are incentivized by interest which accrues on a per block basis.

    \item[Leveraged long position] To take on a long position of an asset refers to purchasing an asset with the expectation that it will appreciate in value. 
    These positions can be taken on a PLF by leveraging the asset on which the long position shall be taken.

    \item[Leveraged short position] A short position refers to borrowing funds of an asset, which one believes will depreciate in value. 
    Consequently, the taker of a short position sells the borrowed asset, only to repurchase it and pay back the lender once the price has fallen, while profiting from the price change of the shorted asset.
    This can be achieved by taking on a leveraged borrow position of the asset to short, with a stablecoin being the locked collateral.

    \item[Liquidity mining] As a means to attract liquidity, PLFs may distribute governance tokens to their liquidity providers.
    The way these tokens are distributed depends on the PLF. 
    For instance, on Compound, the governance token \coin{COMP}\footnote{Contract address: \contractaddr{0xc00e94cb662c3520282e6f5717214004a7f26888}} is distributed among users across markets proportionally to the total dollar value of funds borrowed and supplied.
    This directly incentivizes users to mine liquidity in a market through leveraging in order to receive a larger share of governance tokens.
    For example, a supplier of funds in market $A$ can borrow against his position additional funds of $A$, at the cost of paying the difference between the earned and paid interest. 
    The incentive for pursuing this behaviour exists if the reward (i.e.\ the governance token) exceeds the cost of borrowing.
    
    \item[Token utility] An agent may be able to obtain a token from a PLF which has some desired utility.
    For example, for the case of governance tokens, the desired token utility could be the right to participate in protocol governance or a claim on protocol earnings.

\end{description}

%% file: sections/4_methodology.tex
\section{Methodology}
\label{sec:methodology}

In this section, we describe our methodology for the different analyses we perform with regard to leveraging on a PLF.
To be able to quantify the extent of leveraged positions over time, we first introduce a state transition framework for tracking the supply and borrow positions across all markets on a given PLF.
We then describe how we instantiate this framework on the Compound protocol using on-chain events data.

\subsection{Definitions}

Throughout the paper, we use the following definitions in the context of PLFs:

\begin{description}
\item[Market] A smart contract acting as the intermediary of loanable funds for a particular cryptoasset, where users supply and borrow funds. 

\item[Supply] Funds deposited to a market that can be loaned out to other users and used as collateral against depositors' own borrow positions.

\item[Borrow] Funds loaned out to users of a market.

\item[Collateral] Funds available to back a user's aggregate borrow positions.

\item[Locked funds] Funds remaining in the PLF smart contracts, equal to the difference between supplied and borrowed funds.

\item[Supplier] A user who deposits funds to a market.

\item[Borrower] A user who borrows funds from a market. Since a borrow position must be collateralized by deposited funds, a borrower must also be a supplier.

\item[Liquidator] A user who purchases a borrower's supply in a market when the borrower's collateral-to-borrow ratio falls below some threshold.
\end{description}

\subsection{States on a PLF}
In this section, we provide a formal definition of the state of a PLF.
We denote $\mathfrak{P}_t$ as the global state of a PLF at time $t$.
For brevity, in the following definitions, we assume that all the values are at a given time $t$.
We define the global state for the PLF as
\[
  \mathfrak{P} = (\mathcal{M},\Gamma, \mathcal{P}, \Lambda)
\]
where $\mathcal{M}$ is the set of states of individual markets, $\Gamma$ is the price the Oracle used, $\mathcal{P}$ is the set of states of individual participants and $\Lambda \in (0, 1)$ is the close factor of the protocol, which specifies the upper bound on the amount of collateral a liquidator may purchase.

We define the state of an individual market
$m \in \mathcal{M}$ as 
$$m = (\mathcal{I}, \mathcal{B}, \mathcal{S}, \mathcal{C})$$
where
$\mathcal{I}$ is the market's interest rate model,
$\mathcal{B}$ is the total borrows,
$\mathcal{S}$ is the total supply of deposits,
and $\mathcal{C}$ is the collateral factor.

$\mathcal{P}^m$ is the state of all participants in market $m$ and the positions of a participant $P$ in this market is defined as
\[
  P^m = (B^m,S^m)
\]
where $B^m$ and $S^m$ are respectively the total borrow positions and total supplied deposits of a market participant in market $m$.

For a given market $m$, the total deposits supplied $\mathcal{S}^m$ is thus given by:
\begin{equation}
    \mathcal{S}^m = \sum_{P^m\in \mathcal{P}^m} S^m
\end{equation}

Similarly, the market's total borrows $\mathcal{B}^m$ is given by:
\begin{equation}
    \mathcal{B}^m = \sum_{P^m\in \mathcal{P}^m} B^m
\end{equation}

The state of a participant $P$ is liquidable if the following holds: 

\begin{equation}
    \frac{\sum_{m\in \mathcal{M}}
    \Big\{\left[S^{m} \cdot \mathcal{C} + \mathcal{I}(S^m)\right] \cdot \Gamma(m) \cdot \mathcal{K}^m \Big\}
    }{
    \sum_{m\in \mathcal{M}} \Big\{\left[B^m + \mathcal{I}(B^m)\right] \cdot \Gamma(m)\Big\}}
    < 1
    \label{eq:liqcond}
\end{equation}
where $\Gamma(m)$ returns the price of the underlying asset denominated in a predefined numéraire (e.g. USD), $\mathcal{I}(S^m)$ returns the interest earned with supply $S^m$, $\mathcal{I}(B^m)$ returns the interest accrued with borrow $B^m$, and $\mathcal{K}^m \leq 1$ denotes the liquidation threshold of market $m$. In Compound, liquidation threshold $\mathcal{K}^m$ is set to be constant at 100\% protocol-wide, whereas with other protocols such as Aave, $\mathcal{K}^m$ is specific to the collateral asset from market $m$, and can be dynamically adjusted when the risk level of the asset changes.

The transition from a state of a market $m$ from time $t$ to $t+1$ is given by some state transition $\sigma$, such that $m_{t}\xrightarrow[]{\sigma}m_{t+1}$. 

\subsection{Leveraging Spirals on a PLF}
\label{ssec:leveraging-spirals-meth}
\begin{figure}[tbp]
    \centering
    \includegraphics[width=0.8\textwidth]{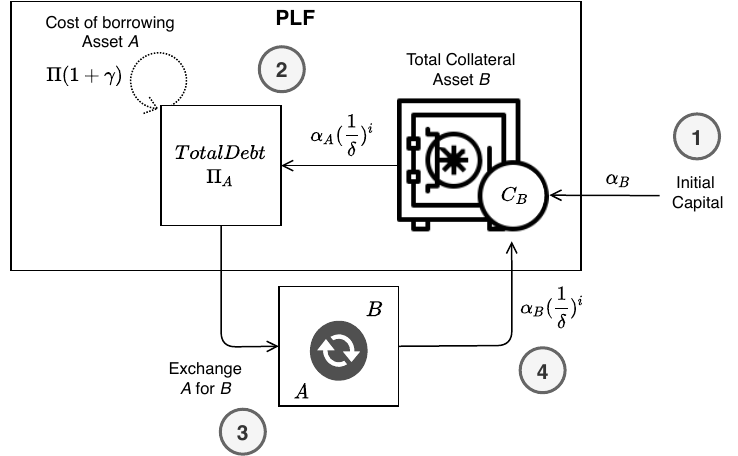}
    \caption{The steps of leveraging using a PLF. \textbf{1.} Initial capital $\alpha_{B}$ in asset $B$ is deposited as collateral to borrow asset $A$. \textbf{2.} Interest accrues over the debt of the borrow position for asset $B$. \textbf{3.} The borrowed asset $A$ is sold for asset $B$ on the open market. \textbf{4.} The newly purchased units of asset $B$ are locked as collateral for a new borrow position of asset $A$.}
    \label{fig:plf-leverage}
\end{figure}

Here we examine the workings of leveraging in DeFi using a PLF. 
We assume a speculator on some volatile asset $B$, holds initial capital $\alpha$ in $B$.
In order to increase his exposure to $B$, the speculator may borrow a stable asset $A$ against his $\alpha$ on a PLF at a collateralization ratio $\delta>1$.
For simplicity, we shall assume in this illustrative example that a speculator will leverage his position on the same PLF.
Note that the cost of borrowing is given by some floating interest rate $\gamma$ for the specific asset market.
In return for his collateral, the borrower receives $\frac{\alpha}{\delta}$ in the volatile asset $B$.
As the debt is denominated in units of a stable asset (e.g. \coin{DAI}), the borrower has an upper limit on his net debt, remaining unaffected by any volatility in the value of asset $A$.
In order to leverage his position, the debt denominated in $A$ may be used to buy\footnote{In practice this may be done via automated market makers \cite{xu2021dexAmm} (e.g. Uniswap~\cite{whitepaper:uniswap}) or via decentralized exchanges~\cite{web:dydx}.} additional units of asset $B$, which can subsequently be used to collateralize a new borrow position.
This process is illustrated in Figure~\ref{fig:plf-leverage} and can be repeated numerous times, by which the total exposure to asset $A$, the underlying collateral to the total debt in asset $A$, increases at a decaying rate.

The total collateral $\mathcal{C}$ a borrower must post through a borrow position with a leverage factor $k$, a collateralization ratio $\delta$ and an initial capital amount $\alpha$ can be expressed as $\sum_{i=0}^{k}\frac{\alpha}{\delta^i}$.
Hence, the total debt $\Pi$ for the corresponding borrow position is:
\begin{equation}
    \Pi = \biggl(\sum_{i=1}^{k} \frac{\alpha}{\delta^i}\biggl) \cdot (1+\gamma)
    \label{eq:debt}
\end{equation}
where $\gamma$ is the interest rate.
Note that Equation~\eqref{eq:debt} assumes a borrower uses the same collateralization ratio $\delta$ for his positions, as well as that all debt is taken out for the same asset on the same PLF and hence the floating interest rate is shared across all borrow positions.

\subsection{States and the Compound PLF}
For our analysis, we apply our state transition framework to the Compound PLF.
Therefore, we briefly present the workings of Compound in the context of our framework.

\input{tables/compound_events_states}

\point{State Transitions}
We initiate state transitions via events emitted from the Compound protocol smart contracts.
We provide an overview of the state variables affected by Compound events in \autoref{tab:compound-events}.

\point{Funds Supplied}
Every market on Compound has an associated ``cToken", a token that continuously appreciates against the underlying asset as interest accrues.
For every deposit in a market, a newly-minted amount of the market's associated cToken is transferred to the depositor.
Therefore, rather than tracking the total amount of the underlying asset supplied, we account the total deposits of an asset supplied by a market participant in the market's cTokens.
Likewise, we account the total supply of deposits in the market in cTokens.

\point{Funds Borrowed}
A borrower on Compound must use cTokens as collateral for his borrow position.
The borrowing capacity equals the current value of the supply multiplied by the collateral factor for the asset.
For example, given an exchange rate of 1~\coin{DAI}~=~50 \coin{cDAI}, a collateral factor of 0.75 for \coin{DAI} and a price of 1~\coin{DAI}~=~1~USD, a holder of 500 \coin{cDAI} (10 \coin{DAI}) would be permitted to borrow up to 7.5 USD worth of some other asset on Compound. 
Therefore, as funds are borrowed, an individual's total borrow position, as well as the respective market's total borrows are updated.

\point{Interest}
The accrual of interest is tracked per market via a borrow index, which corresponds to the total interest accrued in the market.
The borrow index of a market is also used to determine and update the total debt of a borrower in the respective market.
When funds are borrowed, the current borrow index for the market is stored with the borrow position.
When additional funds are borrowed or repaid, the latest borrow index is used to compute the difference of accrued interest since the last borrow and added to the total debt.

\point{Liquidation}
A borrower on Compound is eligible for liquidation should his total supply of collateral, i.e. the value of the sum of the borrower's cToken holdings per market, weighted by each market's collateral factor, be less than the value of the borrower's aggregate debt (Equation~\eqref{eq:liqcond}).
The maximum amount of debt a liquidator may pay back in exchange for collateral is specified by the close factor of a market.

%% file: tables/compound_events_states.tex
\begin{figure}[tb]
  \centering
  \scriptsize
  \setlength{\tabcolsep}{1.5pt}
  \begin{tabular}{lp{6cm}c}
    \toprule
    {\bf Event} & {\bf Description} &
    {\bf State variables affected}\\

    \midrule
    \texttt{Borrow}  & A new borrow position is created. & $\mathcal{B}$  \\ 
    \texttt{Mint} & cTokens are minted for new deposits. &  $\mathcal{S}$           \\
    \texttt{RepayBorrow} & A borrow position is partially/fully repaid.& $\mathcal{B}$            \\ 
    \texttt{LiquidateBorrow} & A borrow position is liquidated. & $\mathcal{B}$, $\mathcal{S}$             \\
    \texttt{Redeem} & cTokens are used to redeem deposits of the underlying asset. & $\mathcal{S}$ \\
    \texttt{NewCollateralFactor} & The collateral factor for the associated market is updated. & $\mathcal{C}$   \\
    \texttt{AccrueInterest} & Interest has accrued for the associated market and its borrow index is updated. & $\mathcal{B}$            \\
    \texttt{NewInterestRateModel} & The interest rate model for the associated market is updated. & $\mathcal{I}$             \\
    \texttt{NewInterestParams} & The parameters of the interest rate model for the associated market are updated. & $\mathcal{I}$            \\
    \texttt{NewCloseFactor} & The close factor is updated. &     $\Lambda$   \\
    \bottomrule
  \end{tabular}
  \caption{The events emitted by the Compound protocol smart contracts used for initiating state transitions and the states affected by each event.}
\label{tab:compound-events}
\end{figure}

%% file: sections/5_analysis.tex
\begin{figure}[tbp]
  \begin{subfigure}{.5\textwidth}
    \centering
    \includegraphics[width=\textwidth]{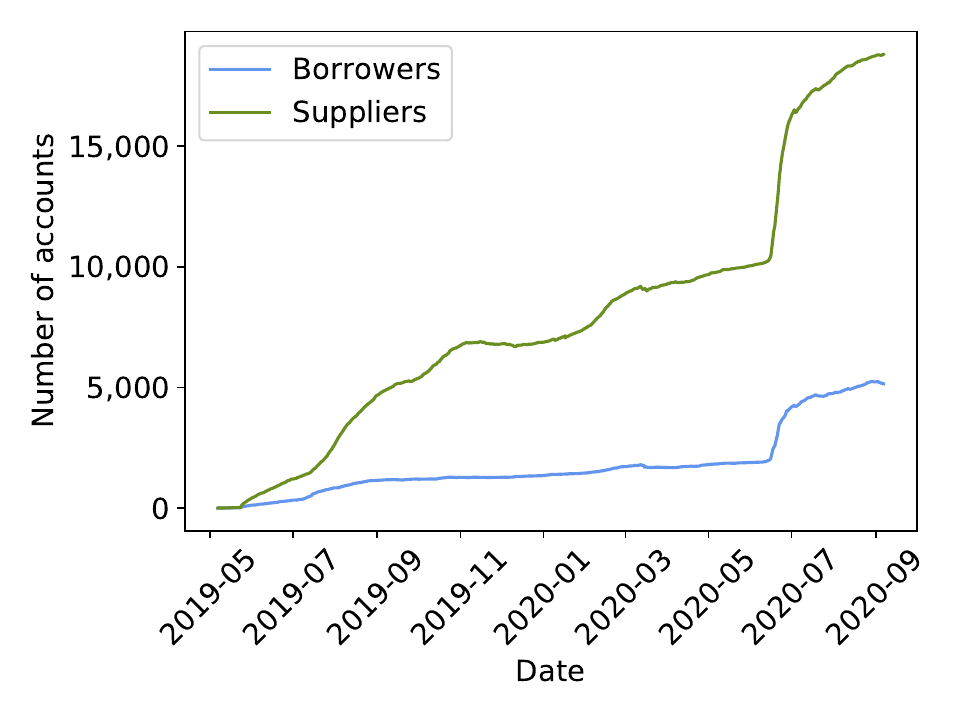}
    \caption{Number of suppliers and borrowers.}
    \label{fig:borrowers-suppliers-over-time}
  \end{subfigure}
  \begin{subfigure}{.5\textwidth}
    \centering
    \includegraphics[width=\textwidth]{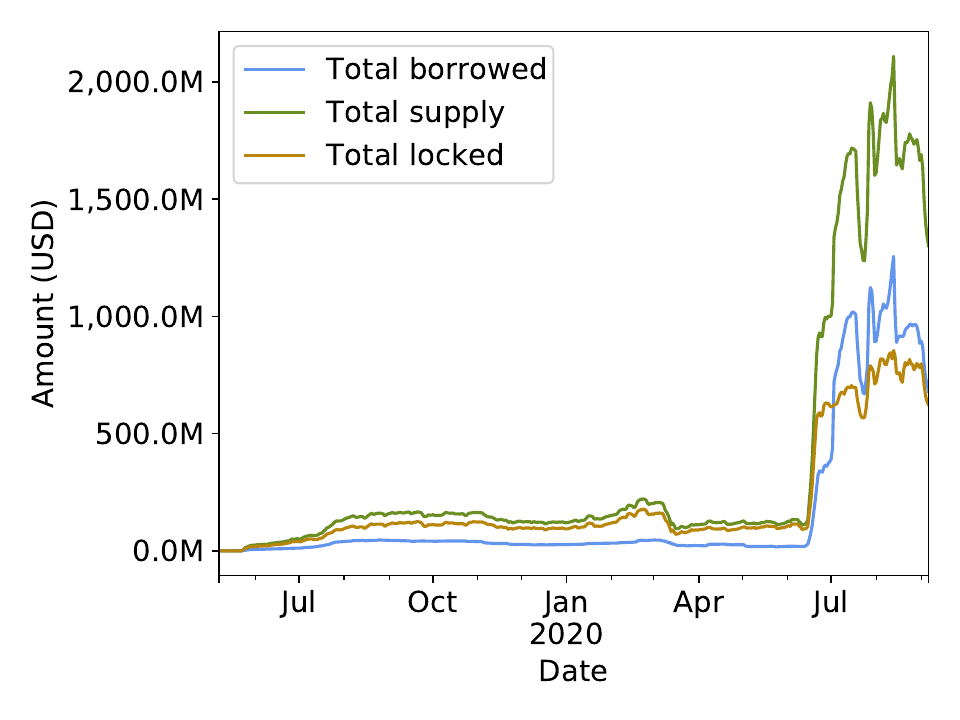}
    \caption{Amount of funds supplied, borrowed and locked.}
    \label{fig:borrow-supply-over-time}
  \end{subfigure}
  \caption{Number of active accounts and amount of funds on Compound over time.}
\end{figure}

\section{Analysis}
\label{sec:analysis}
In this section, we present the results of the analysis performed with the framework outlined in Section~\ref{sec:methodology}.
We analyze data from the Compound protocol~\cite{Leshner2018} over a period ranging from \StartDate---when the first Compound markets were deployed on the Ethereum main network---to \EndDate.
The full list of contracts considered for our analysis can be found in Appendix~\onlineversion{\ref{sec:monitored-contracts}}\printversion{\linktoonline{A}}.
When analyzing a single market, we choose the market for \coin{DAI}, as it is the largest by an order of magnitude.

\begin{figure}[tbp]
  \begin{subfigure}{.5\textwidth}
    \centering
    \includegraphics[width=\textwidth]{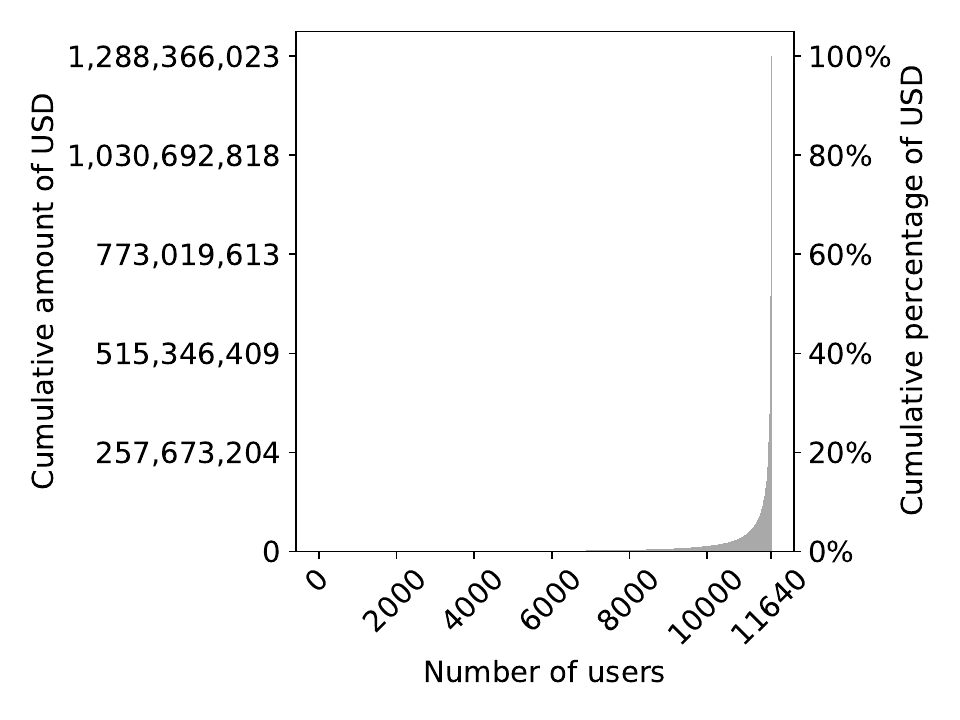}
    \caption{Distribution of supplied funds.}
    \label{fig:suppliers-distribution}
  \end{subfigure}
  \begin{subfigure}{.5\textwidth}
    \centering
    \includegraphics[width=\textwidth]{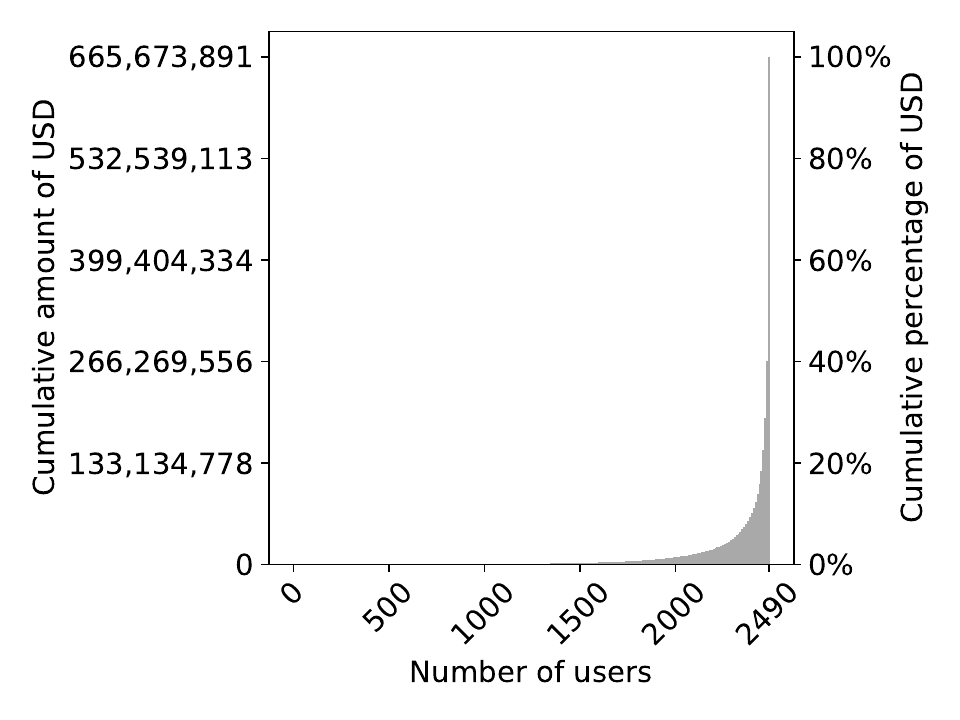}
    \caption{Distribution of borrowed funds.}
    \label{fig:borrowers-distribution}
  \end{subfigure}
  \caption{Cumulative distribution of funds in USD.~Accounts are sorted from least to most wealthy and bucketed in bins of 10, i.e. a single bar represents the sum of 10 accounts.}\label{fig:suppliers-borrowers-distribution}
\end{figure}

\subsection{Borrowers and Suppliers}
We first examine the total number of borrowers and suppliers on Compound by considering any Ethereum account that, at any time within the observation period, either exhibited a non-zero cToken balance or borrowed funds for any Compound market.
The change in the number of borrowers and suppliers over time is displayed in \autoref{fig:borrowers-suppliers-over-time}.

We see that the total number of suppliers always exceeds the total number of borrowers. 
This is because on Compound, one can only borrow against funds he supplied as collateral, which automatically makes the borrower also a supplier. 
Interestingly, the number of suppliers has increased relative to the number of borrowers over time. 
There is notable sudden jump in both the number of suppliers and borrowers in June~2020.

In terms of total deposits, a very similar trend is observable in \autoref{fig:borrow-supply-over-time}, which shows that at the same time, the total supplied deposits increased, while the total borrows followed shortly after.
Furthermore, the total funds borrowed exceeded the total funds locked for the first time in July 2020 and have remained so until the end of the examined period.
We discuss the reasons behind this in the next part of this section.

Despite the similarly increasing trend for the number of suppliers/borrowers and amount of supplied/borrowed funds, we can see in~\autoref{fig:suppliers-borrowers-distribution} that the majority of funds are borrowed and supplied only by a small number of accounts.
For instance, for the suppliers in~\autoref{fig:suppliers-distribution}, the top user and top 10 users supply~27.4\% and~49\% of total funds, respectively.
For the borrowers shown in~\autoref{fig:borrowers-distribution}, the top user accounts for 37.1\%, while the top 10 users account for 59.9\% of total borrows.
While one could think that this concentration comes from the fact that top accounts are pools receiving money from several participants, only one of the top~10 suppliers and none of the top~10 borrowers fit in this category.
We provide a list of the top suppliers and borrowers with a description of the accounts in~\onlineversion{\autoref{tab:top-suppliers-borrowers} of Appendix~\ref{sec:top-suppliers-borrowers}}\printversion{Figure 10 of Appendix \linktoonline{B}}.

\subsection{Leveraging Spirals}
As we have seen in Section~\ref{sec:plfs}, in PLFs, leveraging can be used either to gain more exposure to a particular currency or to gain some incentive provided by the protocol.
To understand how leveraging can affect the total amounts borrowed and supplied on Compound, we use the methodology we defined in Section~\ref{ssec:leveraging-spirals-meth} to measure the existence of leveraging spirals on Compound.

We find that the top supplier deposited a total of 342 million USD and borrowed 247 million.
However, after the inspection of leveraging spirals, we find that the user has provided only 16\% of the funds, while the rest of the minted funds have been part of leveraging spirals, which means that the user provided a total of roughly 55 million USD to the protocol.

In total, we find a total of 2,141 accounts using this leveraging spiral technique for a total of over 600 million USD, or roughly half of the total amount of funds supplied to the protocol.

\subsection{The \coin{COMP} Governance Token}
The sudden jumps exhibited in Figures~\ref{fig:borrowers-suppliers-over-time} and~\ref{fig:borrow-supply-over-time} can be explained by the launch of Compound's governance token, \coin{COMP}, on June~15,~2020.
The \coin{COMP} governance token allows holders to participate in voting, create proposals, as well as delegate voting rights.
In order to empower Compound stakeholders, new \coin{COMP} is minted every block and distributed among borrowers and suppliers in each market.

Initially, \coin{COMP} was allocated proportionally to the accrued interest per market.
However, the \coin{COMP} distribution model was modified via a governance vote on July~2,~2020, such that the borrowing interest rate was removed as a weighting mechanism in favor of distributing \coin{COMP} per market on a borrowing demand basis, i.e.\ per USD borrowed.
The distributed \coin{COMP} per market is shared equally between a market's borrowers and suppliers, who receive \coin{COMP} proportionally to their borrowed and supplied amounts, respectively.
Hence, a Compound user is incentivized to increase his borrow position as long as the borrowing cost does not exceed the value of his \coin{COMP} earnings. This presumably explains the drop in the degree of collateralization, as the total amount locked is seen surpassed by the total borrows after the \coin{COMP} launch (\autoref{fig:borrow-supply-over-time}), leading to elevated liquidation risk of borrow positions.

\begin{figure}[tbp]
  \centering
  \includegraphics[width=0.7\textwidth]{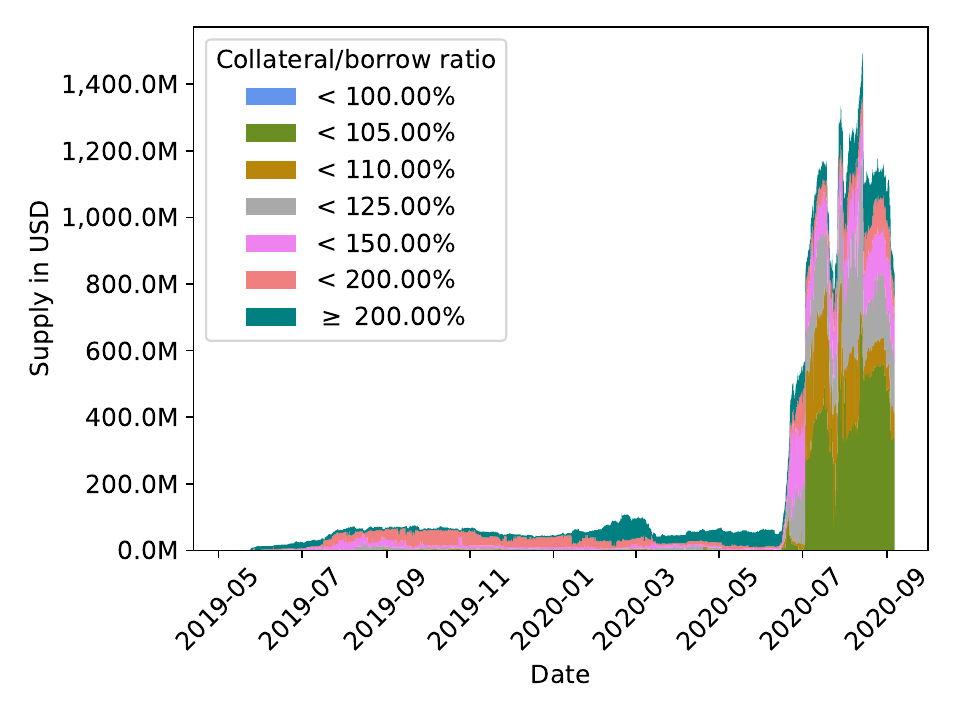}
  \caption{Collateral locked over time, showing how close the amounts are from being liquidated. Positions can be liquidated when the ratio drops below~100\%.}
  \label{fig:liquidatability}
\end{figure}

\subsection{Liquidation Risk}
Given the high increase in the number of total funds borrowed and supplied, as well as the decrease in liquidity relative to total borrows, we seek to identify and quantify any changes in liquidation risk on Compound since the launch of \coin{COMP}. 
\autoref{fig:liquidatability} shows the total USD value of collateral on Compound and how close collateral amounts are from liquidation. 
In addition to the substantial increase in the total value of collateral on Compound since the launch of \coin{COMP}, the risk-seeking behavior of users has also changed.
This can be seen by examining collateral to borrow ratios, where since beginning of July,~2020, a total of approximately 350m to 600m USD worth of collateral has been within a 5\% price range of becoming liquidable.
However, it should be noted that the likelihood of the amount of this collateral becoming liquidable highly depends on the price volatility of the collateral asset.

\begin{figure}[tb]
  \centering
  \includegraphics[width=.7\textwidth]{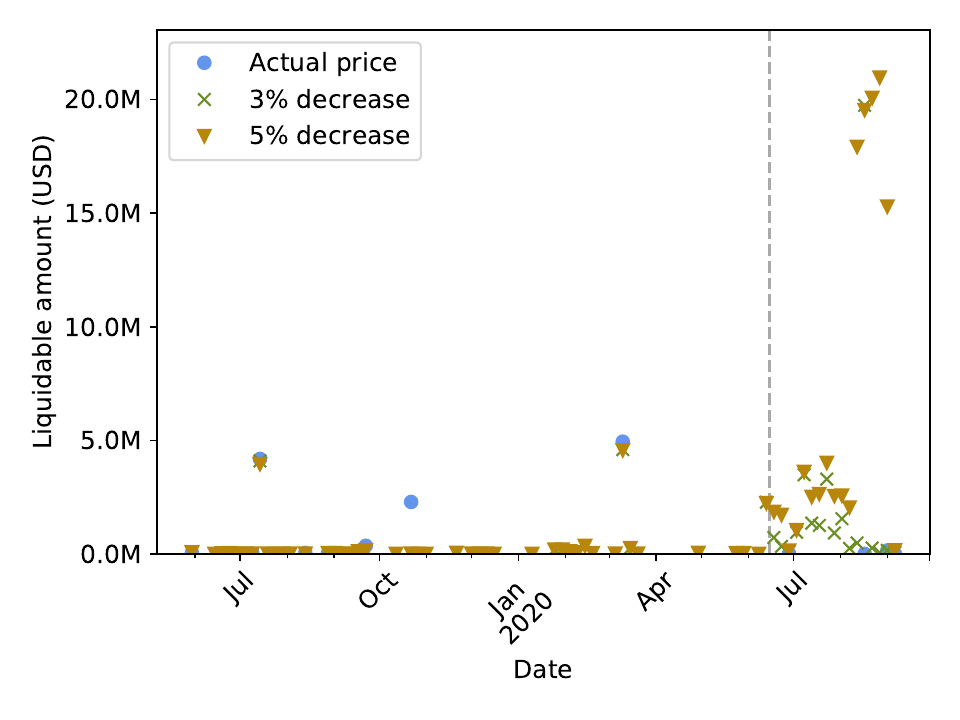}
  \caption{Sensitivity analysis of the liquidable collateral amount given \coin{DAI} price movement relative to its peg USD. \coin{COMP} launch date is marked by the dashed vertical line.}
  \label{fig:price-sensitivity}
\end{figure}

In order to examine how liquidation risk differs across markets, we measure for the largest market on Compound, namely \coin{DAI}, the sensitivity of collateral becoming liquidable given a decrease in the price of \coin{DAI}.
\autoref{fig:price-sensitivity} shows the amount of aggregate collateral liquidable at the historic price, as well as at a~3\% and~5\% decrease relative to the historic price for \coin{DAI}.
We mark the date on which the \coin{COMP} governance token launched with a dashed line.
It can be seen that since the launch of \coin{COMP},~3\% and~5\% price decreases of \coin{DAI} relative to its peg USD would have resulted in a substantially higher amount of liquidable collateral.
In particular, a~3\% decrease would have turned collateral worth in excess of~10 million USD liquidable.

\begin{figure}[tb]
  \centering
  \includegraphics[width=.7\textwidth]{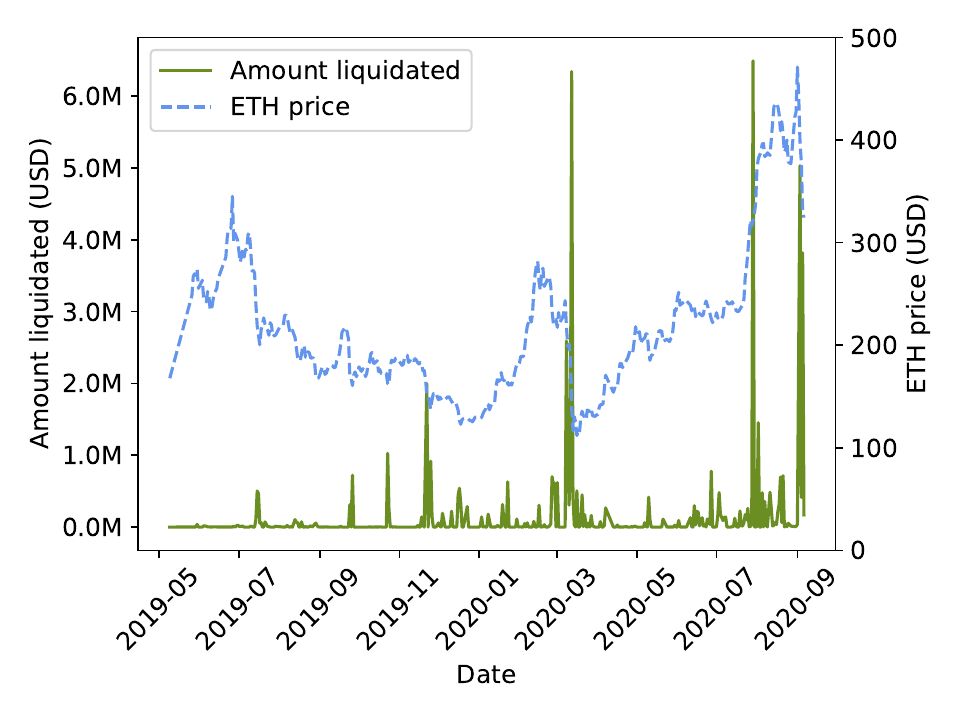}
  \caption{Amount (in USD) of liquidated collateral from May ~2019 to August~2020.}
  \label{fig:liquidations-over-time}
\end{figure}

\subsection{Liquidations and Liquidators}
In order to better understand the implications of the increased liquidation risk since the launch of \coin{COMP}, we examine historical liquidations on Compound and subsequently measure the efficiency of liquidators.

\point{Historical Liquidations}
The increased risk-seeking behavior suggested by the low collateral to borrow ratios presented in the previous section are in accordance with the trend of rising amount of liquidated collateral since the introduction of \coin{COMP}.  
The total value of collateral liquidated on Compound over time is shown in \autoref{fig:liquidations-over-time}.
It can be seen that the majority of this collateral was liquidated on a few occasions, perhaps most notably on Black Thursday (March 12, 2020), July 29, 2020 (\coin{DAI} deviating from its peg significantly), and in early September 2020 (\coin{ETH} price drop).

\begin{figure}[tbp]
    \centering
    \includegraphics[width=0.7\textwidth]{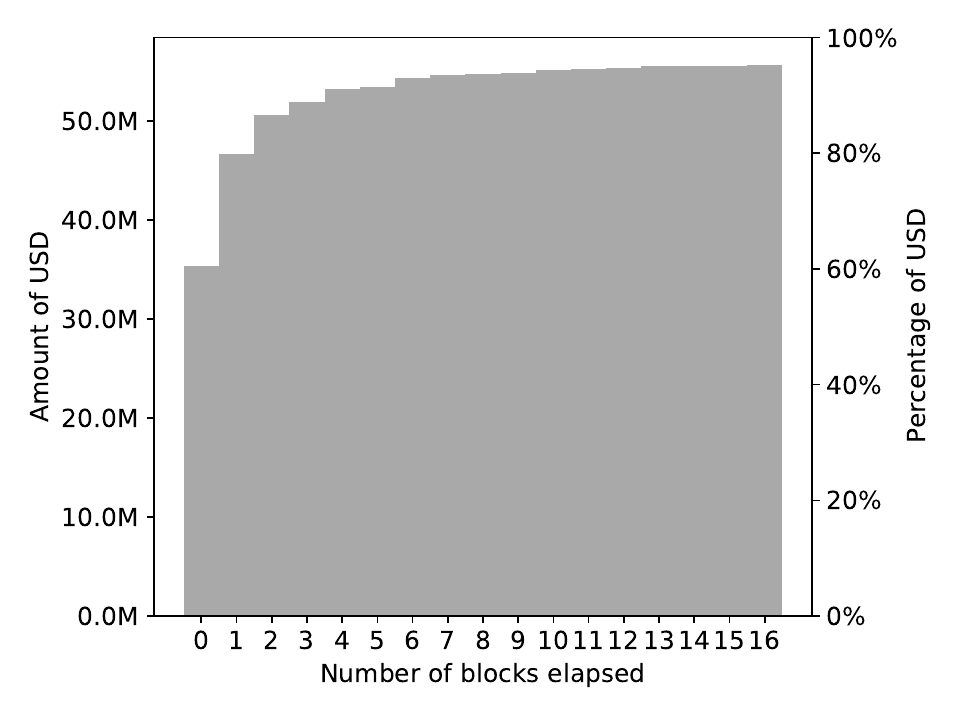}
    \caption{Number of blocks elapsed from the time a position can be liquidated to actual liquidation on Compound from \StartDate to \EndDate, shown as a CDF.}
    \label{fig:blocks-spent}
\end{figure}

\point{Liquidation Efficiency}
We measure the efficiency of liquidators as the number of blocks elapsed since a borrow position has become liquidable and the position actually being liquidated.
The overall historical efficiency of liquidators is shown as a cumulative distribution function in \autoref{fig:blocks-spent}, from which it can be seen that approximately~60\% of the total liquidated collateral (35 million USD) was liquidated within the same block as it became liquidable, suggesting that the majority of liquidations occur via bots in a highly efficient fashion.
After~2 blocks have elapsed (on average half a minute),~85\% of liquidable collateral has been liquidated, and after 16 blocks this value amounts to~95\%.

It is worth noting that liquidation efficiency has been skewed by the more recent liquidation activities which were of a much larger scale than when the protocol was first launched.
Specifically, in 2019, only about~26\% of the liquidations occurred in the block during which the position became liquidable, compared to~70\% in 2020.
This resulted in some lost opportunities for liquidators as shown in~\autoref{fig:price-sensitivity}.
The account \contractaddr{0xd062eeb318295a09d4262135ef0092979552afe6}, for instance, had more than 3,000,000 USD worth of \coin{ETH} as collateral exposed at block 8,796,900 for the duration of a single block: the account was roughly 20 USD shy of the liquidation threshold but eventually escaped liquidation.
If a liquidator had captured this opportunity, he could have bought half of this collateral (given the close factor of 0.5), at a 10\% discount, resulting in a profit of 150,000 USD for a single transaction.
It is clear that with such stakes, participants were incentivized to improve liquidation techniques, resulting in a high level of liquidation speed and scale.

\subsection{Summary}
In this section, we have analyzed the Compound protocol with a focus on liquidations.
We have found that despite the increase in number of suppliers and borrowers over time, the total amount of funds supplied and borrowed remain extremely concentrated among a small set of participants.

We have also seen that the introduction of the \coin{COMP} governance token has changed how users interact with the protocol and the amount of risk that they are willing to take.
Users now borrow vastly more than before, with the total amount borrowed surpassing the total amount locked.
Due to excessive borrowing without a sufficiently safe amount of supplied funds, borrow positions now face a higher liquidation risk, such that a crash of~3\% in the price of \coin{DAI} could result in an aggregate liquidation value of over~10 million USD.

Finally, we have shown that the liquidators have become more efficient with time, and are currently able to capture a majority of the liquidable funds instantly.

%% file: sections/7_discussion.tex
\section{Discussion}
\label{sec:discussion}
In this section, we enumerate several points that we deem important for the future development of PLFs and DeFi protocols.
We first discuss the influence of governance tokens, by intention or not, on how users behave within a protocol.
Subsequently, we discuss potential risks that lie in the use of governance tokens, and the contagion effect that user behavior in a protocol can have on another protocol.
Finally, we discuss how miner-extractable value~\cite{daian2020flash} can potentially affect liquidation incentives in such protocols.

\subsection{Governance Token Influence}
As analyzed in Section~\ref{sec:analysis}, the distribution of the \coin{COMP} token has vastly changed the Compound landscape and user behavior.
Until the introduction of the token, borrowing was costly due to the payable interest, which implies a negative cash flow for the borrower. Therefore, a borrower would only borrow if he could justify this negative cash flow with some application external to Compound.
With the introduction of this token, borrowing could yield a positive cash flow due to the monetary value of the governance token.
This creates a situation where both suppliers and borrowers end up with a positive cash flow, inducing users to maximize both their supply and borrow.
This model is, however, only sustainable when the price of the \coin{COMP} token remains sufficiently high to keep this cash flow positive for borrowers.
This directly results in users taking increasingly higher risk in an attempt to gain larger monetary rewards, with liquidators ultimately profiting more from their operations.

\subsection{Governance Token Risks}
The increased use of governance tokens across DeFi protocols (e.g. \coin{YFI} on Yearn Finance, \coin{AAVE} on Aave, \coin{UNI} on Uniswap) can be seen as a promising step towards achieving a higher degree of decentralization in terms of protocol governance. 
However, despite the increased usage of governance tokens, to the best of our knowledge there is still a dearth of academic research examining the different governance models and specifically the relation between their security assumptions and the employed governance token.
For instance, the option to aggregate governance tokens via flash loans \cite{Wang2020} can pose a significant security risk to DeFi protocols should an attacker attempt to propose and execute malicious protocol updates.
Furthermore, even in the case of flash loan resistant governance models, the relationship between the financial value of a protocol's governance token and the economically secure regions of the protocol remains unexamined and serves as a further risk that designers of governance models have to take into account.
Despite the existence of protective mechanisms against governance attacks on some protocols (e.g. multi-sig approvals or selected ``guardians'' that are able to halt the governance process), it remains questionable which of such mechanisms are indeed desirable from a decentralized governance perspective and whether there might be more suitable alternatives.

\subsection{Contagion Effects}
This behavior also indirectly affected other protocols, in particular \coin{DAI}.
The price of \coin{DAI} is aimed to be pegged to 1 USD resting on an arbitrage mechanism, whereby token holders are incentivized to buy or sell \coin{DAI} as soon as the price moves below or above 1 USD, respectively.
However, a rational user seeking to maximize profit will not sell his \coin{DAI} if holding it somewhere else would yield higher profits.
This was precisely what was happening with Compound, whose users locking their \coin{DAI} received higher yields in the form of \coin{COMP}, than from selling \coin{DAI} at a premium, thereby resulting in upward price pressure~\cite{cyrus2020upcoming}.
Interestingly, \coin{DAI} deviating from its peg also has a negative effect for Compound users.
Indeed, as we saw in Section~\ref{sec:analysis}, many Compound users might have been overconfident about the price stability of \coin{DAI} and thus only collateralize marginally above the threshold when they borrow \coin{DAI}.
This has resulted in large amounts being liquidated due to the actual, higher extent of the volatility in the \coin{DAI} price.

\subsection{Miner-Extractable Value}
In the context of PLFs, liquidations can be seen as miner-extractable value.
Indeed, it is easy for the miner to check whether a position is liquidable or not after each processed transaction and to add a transaction to liquidate the position immediately after the transaction making it liquidable.
In our analysis of the Compound protocol, we have not found any sign of miners participating in liquidations, directly or indirectly.
We show the top miners and the top liquidators for each miner in \onlineversion{\autoref{tab:miners-liquidators} of Appendix~\ref{sec:miners-liquidators}}\printversion{Figure 11 of Appendix \linktoonline{C}}.
Although we found no correlation between miners and liquidators, this is a real risk that could make the role of liquidator, which is essential for the protocol security, less interesting for those who are not collaborating with miners.

%% file: sections/8_related_work.tex
\section{Related Work}
\label{sec:related-work}

In this section we briefly discuss existing work related to this paper.

A thorough analysis of the Compound protocol with respect to market risks faced by participants was done by~\cite{Kao2020}.
The authors employ agent-based modeling and simulation to perform stress tests in order to show that Compound remains safe under high volatility scenarios and high levels of outstanding debt.
Furthermore, the authors demonstrate the potential of Compound to scale to accommodate a larger borrow market while maintaining a low default probability.
This differs from our work as we conduct a detailed empirical analysis on Compound, focusing on how agent behavior under different incentive structures on Compound has affected the protocol's state with regard to liquidation risk.  

A first in-depth analysis on PLFs is given by~\cite{Gudgeon2020PLF}.
The authors provide a taxonomy on interest rate models employed by PLFs, while also discussing market liquidity, efficiency and interconnectedness across PLFs.
As part of their analysis, the authors examine the cumulative percentage of locked funds solely for the Compound markets \coin{DAI}, \coin{ETH}, and \coin{USDC}.

In~\cite{bartoletti2020sok}, the authors provide a formal state transition model of PLFs\footnote{Note that in \cite{bartoletti2020sok}, PLFs are referred to as lending pools.} and prove fundamental behavioural properties of PLFs, which had previously only been presented informally in the literature.
Additionally, the authors examine attack vectors and risks, such as utilization attacks and interest bearing derivative token risk. 
This work differs to our work, as the authors of~\cite{bartoletti2020sok} formalize the properties of PLFs through an abstract model, while we provide a thorough empirical analysis with a focus on liquidations and risks brought upon by governance tokens, such as for Compound and the \coin{COMP} token.

In \cite{klages2019stability}, the authors show how markets for stablecoins are exposed to deleveraging feedback effects, which can cause periods of illiquidity during crisis.

The authors of \cite{gudgeon2020decentralized} demonstrate how various DeFi lending protocols are subject to different attack vectors such as governance attacks and undercollateralization.
In the context of the proposed governance attack, the lending protocol the authors focus on is Maker~\cite{whitepaper:maker}.





%% file: sections/9_conclusion.tex
\section{Conclusion}
\label{sec:conclusion}
In this paper, we presented the first in-depth empirical analysis of liquidations on Compound, one of the largest PLFs in terms of total locked funds, from~\StartDate to~\EndDate.
We analyzed agents' behavior and in particular how much risk they are willing to take within the protocol.
Furthermore, we assessed how this has changed with the launch of the Compound governance token \coin{COMP}, where we found that agents take notably higher risks in anticipation of higher earnings.
This resulted in variations as little as 3\% in an asset's price being able to turn over 10 million USD worth of collateral liquidable.
In order to better understand the potential consequences, we then measured the efficiency of liquidators, namely how quickly new liquidation opportunities are captured. Liquidators' efficiency was found to have improved significantly over time, reaching 70\% of instant liquidations.
Lastly, we demonstrated how overconfidence in the price stability of \coin{DAI}, increased the overall liquidation risk faced by Compound users.
Rather ironically, many users wishing to make the most of the new incentive scheme ended up causing higher volatility in \coin{DAI}---a dominant asset of the platform, resulting in liquidation of their own assets.
This is not Compound's misdoing, but rather highlights the to date unknown dynamics of incentive structures across different DeFi protocols.

%% file: sections/a_contracts.tex
\section{Monitored Contracts}
\label{sec:monitored-contracts}
In \autoref{fig:monitored-contracts}, we provide a list of contracts we monitored in our analysis.

\begin{figure}[h!]
  \centering
  \setlength{\tabcolsep}{5pt}
  \begin{tabular}{l l}
    \toprule
    Name & Address\\
    \midrule
    cBAT & \contractaddr{0x6c8c6b02e7b2be14d4fa6022dfd6d75921d90e4e}\\
    cDAI & \contractaddr{0x5d3a536e4d6dbd6114cc1ead35777bab948e3643}\\
    cETH & \contractaddr{0x4ddc2d193948926d02f9b1fe9e1daa0718270ed5}\\
    cREP & \contractaddr{0x158079ee67fce2f58472a96584a73c7ab9ac95c1}\\
    cSAI & \contractaddr{0xf5dce57282a584d2746faf1593d3121fcac444dc}\\
    cUSDC & \contractaddr{0x39aa39c021dfbae8fac545936693ac917d5e7563}\\
    cUSDT & \contractaddr{0xf650c3d88d12db855b8bf7d11be6c55a4e07dcc9}\\
    cWBTC & \contractaddr{0xc11b1268c1a384e55c48c2391d8d480264a3a7f4}\\
    cZRX & \contractaddr{0xb3319f5d18bc0d84dd1b4825dcde5d5f7266d407}\\
    Comptroller & \contractaddr{0x3d9819210a31b4961b30ef54be2aed79b9c9cd3b}\\
    Open Oracle Price Data & \contractaddr{0x02557a5e05defeffd4cae6d83ea3d173b272c904}\\
    Uniswap Anchored View & \contractaddr{0x9b8eb8b3d6e2e0db36f41455185fef7049a35cae}\\
    \bottomrule
  \end{tabular}
  \caption{Monitored contracts}
  \label{fig:monitored-contracts}
\end{figure}

%% file: sections/b_extras.tex
\section{Top Suppliers and Borrowers}
\label{sec:top-suppliers-borrowers}
In~\autoref{tab:top-suppliers-borrowers}, we show the top 10 suppliers and borrowers in terms of amount borrowed and supplied expressed in USD.
We mark the addresses that are smart contracts. Among those contracts, the one with the most supplied funds, \contractaddr[\scriptsize]{0xa2b47e3d5c44877cca798226b7b8118f9bfb7a56}, is the address of a Curve~\cite{web:curve} pool that has funds coming from a several independent parties.
No other address among top suppliers and borrowers is a pool address.

\begin{figure}[tbp]
  \setlength{\tabcolsep}{3.5pt}
  \begin{subfigure}{\textwidth}
    \begin{tabular}{lcrl}
      \toprule
      \textbf{Address} &  & \textbf{Amount} & \textbf{Description}\\
      \midrule
      \contractaddr[\scriptsize]{0x554bd2947df1c8d8d38897bdc92b3b97692b2845} &   & 342,128,032 &\\
      \contractaddr[\scriptsize]{0xa2b47e3d5c44877cca798226b7b8118f9bfb7a56} & \checkmark  & 40,284,236 & Curve pool\\
      \contractaddr[\scriptsize]{0x04b0b0e460c9fc583d9c93bc9ae25b353390645e} & \checkmark  & 34,908,472 & Instadapp smart wallet\\
      \contractaddr[\scriptsize]{0x25599dcbd434af9a17d52444f71c92987fa97cfc} &  & 34,530,570 & \\
      \contractaddr[\scriptsize]{0x58485ea7106891bdd94c37ced30c6fdbc5293b16} & \checkmark  & 32,686,029 & Multisig wallet\\
      \contractaddr[\scriptsize]{0x909b443761bbd7fbb876ecde71a37e1433f6af6f} &  & 29,308,425  & \\
      \contractaddr[\scriptsize]{0xea61f3052753ea2c6a1c208583ad9b0394ed2f28} & \checkmark  & 28,854,366 & DeFi Saver smart wallet\\
      \contractaddr[\scriptsize]{0x32b2d4ec46d76fc6dabfe958fb0e0bd8db740c84} &  & 27,928,637 & \\
      \contractaddr[\scriptsize]{0xedcc13d25e23032b61d30c298334f92d7c0ba84e} &  & 27,709,153 & \\
      \contractaddr[\scriptsize]{0x6d2af065ccb60c0f7e8ec5907c961c42a3447127} &  & 25,559,037 & \\
      \bottomrule
    \end{tabular}
    \caption{Top 10 users with the largest amount of funds supplied}
  \end{subfigure}
  \begin{subfigure}{\textwidth}
    \begin{tabular}{llrl}
      \toprule
      \textbf{Address} &  & \textbf{Amount} & \textbf{Description}\\
      \midrule
      \contractaddr[\scriptsize]{0x554bd2947df1c8d8d38897bdc92b3b97692b2845} & & 247,143,532 & \\
      \contractaddr[\scriptsize]{0x25599dcbd434af9a17d52444f71c92987fa97cfc} & & 22,085,613 & \\
      \contractaddr[\scriptsize]{0x909b443761bbd7fbb876ecde71a37e1433f6af6f} & & 21,030,095 & \\
      \contractaddr[\scriptsize]{0x58485ea7106891bdd94c37ced30c6fdbc5293b16} & \checkmark & 20,149,687 & Multisig wallet\\
      \contractaddr[\scriptsize]{0x32b2d4ec46d76fc6dabfe958fb0e0bd8db740c84} & & 18,900,729 & \\
      \contractaddr[\scriptsize]{0xea61f3052753ea2c6a1c208583ad9b0394ed2f28} & \checkmark & 18,248,324 & DeFi Saver smart wallet\\
      \contractaddr[\scriptsize]{0xedcc13d25e23032b61d30c298334f92d7c0ba84e} & & 17,643,172 & \\
      \contractaddr[\scriptsize]{0x6d2af065ccb60c0f7e8ec5907c961c42a3447127} & & 12,015,576 & \\
      \contractaddr[\scriptsize]{0x79dbd1baf124edd4205b2aba56c29bf3914c8ed0} & & 11,632,820 & \\
      \contractaddr[\scriptsize]{0x0c8a8dd439069690a5722d5fbb18359a68e279f1} & & 10,009,553 & \\
      \bottomrule
    \end{tabular}
    \caption{Top 10 users with the largest amount of funds borrowed}
  \end{subfigure}
  \caption{Top 10 suppliers and borrowers. Amounts are expressed in their USD equivalent. Addresses marked with $\checkmark$ are smart contract addresses, among which the one with the most supplied funds is a Curve pool address that aggregates funds from multiple parties. 
  }
  \label{tab:top-suppliers-borrowers}
\end{figure}

\section{Relationship Between Miners and Liquidators}
\label{sec:miners-liquidators}
In~\autoref{tab:miners-liquidators}, we show the 10 miners who mined the most blocks containing at least one liquidation.
For each miner, we show the 5 liquidators who liquidated the most positions in blocks mined by the given miner.
Overall, we see that for every miner, the liquidations are spread relatively evenly across the different liquidators.
Although we only show the top 10 miners for space constraints, we noted that this was the case for all miners in our dataset.

\hypersetup{hidelinks}
\renewcommand{\contractaddr}[2][\ssmall]{{#1\href{https://etherscan.io/token/#2}{\texttt{#2}}}}
\begin{figure}
  \scriptsize
  \renewcommand{\arraystretch}{1.2}
  \begin{tabular}{@{}l@{}rl@{}r@{}}
    \toprule
    \textbf{Miner} & \textbf{Blocks} & \textbf{Liquidators} & \textbf{Liquidations} \\
          & \textbf{count} & & \textbf{count}\\
    \midrule
\multirow{5}{*}{\contractaddr{0x5A0b54D5dc17e0AadC383d2db43B0a0D3E029c4c}} & \multirow{5}{*}{1281} & \contractaddr{0x6a0c50788E462f322959A2458687096994d66316} & 144\\
                                           &       & \contractaddr{0x8c863333c2E92f02e01F7A3c6d131E4d59f78990} & 114\\
                                           &       & \contractaddr{0x0c31b6605686aa26df47eb45AF0e4aa6639A5fd6} &  91\\
                                           &       & \contractaddr{0xb00ba6778cF84100da676101e011B3d229458270} &  76\\
                                           &       & \contractaddr{0x268a1b7ECC1fE1FaB1eE32a7e61e3b7810BAD4a5} &  70\\
\hline
\multirow{5}{*}{\contractaddr{0xEA674fdDe714fd979de3EdF0F56AA9716B898ec8}} & \multirow{5}{*}{969} & \contractaddr{0x6a0c50788E462f322959A2458687096994d66316} &  88\\
                                           &       & \contractaddr{0xb00ba6778cF84100da676101e011B3d229458270} &  75\\
                                           &       & \contractaddr{0x8c863333c2E92f02e01F7A3c6d131E4d59f78990} &  70\\
                                           &       & \contractaddr{0x0c31b6605686aa26df47eb45AF0e4aa6639A5fd6} &  52\\
                                           &       & \contractaddr{0x268a1b7ECC1fE1FaB1eE32a7e61e3b7810BAD4a5} &  50\\
\hline
\multirow{5}{*}{\contractaddr{0x829BD824B016326A401d083B33D092293333A830}} & \multirow{5}{*}{310} & \contractaddr{0x6a0c50788E462f322959A2458687096994d66316} &  31\\
                                           &       & \contractaddr{0x8c863333c2E92f02e01F7A3c6d131E4d59f78990} &  26\\
                                           &       & \contractaddr{0x402a75f3500CA1FbA17741Ec916F07a0c9DB195D} &  23\\
                                           &       & \contractaddr{0xb00ba6778cF84100da676101e011B3d229458270} &  18\\
                                           &       & \contractaddr{0x029720A9b3CE72f3e1D9C79257E1F19AfE20b6c9} &  17\\
\hline
\multirow{5}{*}{\contractaddr{0x52bc44d5378309EE2abF1539BF71dE1b7d7bE3b5}} & \multirow{5}{*}{257} & \contractaddr{0x6a0c50788E462f322959A2458687096994d66316} &  22\\
                                           &       & \contractaddr{0x10aab4B0EF76AA2AC9b5909e671517a1171B050E} &  21\\
                                           &       & \contractaddr{0x8c863333c2E92f02e01F7A3c6d131E4d59f78990} &  16\\
                                           &       & \contractaddr{0x402a75f3500CA1FbA17741Ec916F07a0c9DB195D} &  15\\
                                           &       & \contractaddr{0x0006e4548AED4502ec8c844567840Ce6eF1013f5} &  14\\
\hline
\multirow{5}{*}{\contractaddr{0x04668Ec2f57cC15c381b461B9fEDaB5D451c8F7F}} & \multirow{5}{*}{185} & \contractaddr{0x8c863333c2E92f02e01F7A3c6d131E4d59f78990} &  22\\
                                           &       & \contractaddr{0x5DAfafbd7AcD662C909a9601120cf1D9F277e8aE} &  14\\
                                           &       & \contractaddr{0x10aab4B0EF76AA2AC9b5909e671517a1171B050E} &  14\\
                                           &       & \contractaddr{0x268a1b7ECC1fE1FaB1eE32a7e61e3b7810BAD4a5} &  12\\
                                           &       & \contractaddr{0x6a0c50788E462f322959A2458687096994d66316} &  12\\
\hline
\multirow{5}{*}{\contractaddr{0xb2930B35844a230f00E51431aCAe96Fe543a0347}} & \multirow{5}{*}{77} & \contractaddr{0x10aab4B0EF76AA2AC9b5909e671517a1171B050E} &   8\\
                                           &       & \contractaddr{0x0c31b6605686aa26df47eb45AF0e4aa6639A5fd6} &   8\\
                                           &       & \contractaddr{0x5DAfafbd7AcD662C909a9601120cf1D9F277e8aE} &   6\\
                                           &       & \contractaddr{0xf8E562f4F30c5DdA0978857067D6585265dA3437} &   6\\
                                           &       & \contractaddr{0xfDe817C7a0770f42fb80B93dd7A538291C871765} &   5\\
\hline
\multirow{5}{*}{\contractaddr{0xD224cA0c819e8E97ba0136B3b95ceFf503B79f53}} & \multirow{5}{*}{73} & \contractaddr{0xb00ba6778cF84100da676101e011B3d229458270} &  13\\
                                           &       & \contractaddr{0x8c863333c2E92f02e01F7A3c6d131E4d59f78990} &   8\\
                                           &       & \contractaddr{0x88886841CfCCBf54AdBbC0B6C9cBAceAbec42b8B} &   8\\
                                           &       & \contractaddr{0xfFA7370a03c2a91f5B1847a90750489d05f52Fa9} &   5\\
                                           &       & \contractaddr{0x492Ff1c96b398297FcAcd6E7E1E968d2b2fc7Da0} &   5\\
\hline
\multirow{5}{*}{\contractaddr{0x4C549990A7eF3FEA8784406c1EECc98bF4211fA5}} & \multirow{5}{*}{68} & \contractaddr{0xb00ba6778cF84100da676101e011B3d229458270} &  12\\
                                           &       & \contractaddr{0x6a0c50788E462f322959A2458687096994d66316} &   9\\
                                           &       & \contractaddr{0x402a75f3500CA1FbA17741Ec916F07a0c9DB195D} &   5\\
                                           &       & \contractaddr{0x10aab4B0EF76AA2AC9b5909e671517a1171B050E} &   4\\
                                           &       & \contractaddr{0x8c863333c2E92f02e01F7A3c6d131E4d59f78990} &   3\\
\hline
\multirow{5}{*}{\contractaddr{0xEEa5B82B61424dF8020f5feDD81767f2d0D25Bfb}} & \multirow{5}{*}{55} & \contractaddr{0xb00ba6778cF84100da676101e011B3d229458270} &   7\\
                                           &       & \contractaddr{0x402a75f3500CA1FbA17741Ec916F07a0c9DB195D} &   6\\
                                           &       & \contractaddr{0x8c863333c2E92f02e01F7A3c6d131E4d59f78990} &   5\\
                                           &       & \contractaddr{0x029720A9b3CE72f3e1D9C79257E1F19AfE20b6c9} &   5\\
                                           &       & \contractaddr{0x10aab4B0EF76AA2AC9b5909e671517a1171B050E} &   4\\
\hline
\multirow{5}{*}{\contractaddr{0x84A0d77c693aDAbE0ebc48F88b3fFFF010577051}} & \multirow{5}{*}{46} & \contractaddr{0xb00ba6778cF84100da676101e011B3d229458270} &   7\\
                                           &       & \contractaddr{0x0c31b6605686aa26df47eb45AF0e4aa6639A5fd6} &   6\\
                                           &       & \contractaddr{0x6a0c50788E462f322959A2458687096994d66316} &   5\\
                                           &       & \contractaddr{0x5e32f33e261a90FF9fE94230387118945599268c} &   5\\
                                           &       & \contractaddr{0x8c863333c2E92f02e01F7A3c6d131E4d59f78990} &   5\\
    \bottomrule
  \end{tabular}
  \caption{Top 10 miners per number of blocks containing at least a liquidation event mined and top 5 liquidators for each miner per number of liquidations}\label{tab:miners-liquidators}
\end{figure}
\hypersetup{pdfborder={0 0 1}}